\documentclass[12pt,a4paper]{article}
\usepackage[british]{babel}
\usepackage{epsfig}
                                                                                
\begin{document}                                                                
\date{}

\title{ Dynamical fermion algorithm for variable actions }

\author{ I.\ Montvay                             \\
         Deutsches Elektronen-Synchrotron DESY   \\
         Notkestr.\,85, D-22603 Hamburg, Germany }

\newcommand{\be}{\begin{equation}}                                              
\newcommand{\ee}{\end{equation}}                                                
\newcommand{\half}{\frac{1}{2}}                                                 
\newcommand{\rar}{\rightarrow}                                                  
\newcommand{\lar}{\leftarrow}
\newcommand{\LCB}{\raisebox{-0.3ex}{\mbox{\LARGE$\left\{\right.$}}}
\newcommand{\RCB}{\raisebox{-0.3ex}{\mbox{\LARGE$\left.\right\}$}}}
\newcommand{\U}{\mathrm{U}}
\newcommand{\SU}{\mathrm{SU}}
                                                                                
\maketitle
\vspace*{1em}

\begin{abstract} \normalsize
 A new version of the two-step multi-boson algorithm is developed
 with different fermion actions in the multi-boson and noisy
 Metropolis steps.
\end{abstract}       

\section{Introduction}\label{sec1}

 Multi-boson algorithms for dynamical fermions
 \cite{MULTIB,CORR,TSMB,PHMC} offer promising possibilities for
 numerical simulations of QCD and other similar theories \cite{PISA}.
 The two-step multi-boson (TSMB) algorithm \cite{TSMB} has been
 succesfully applied, for instance, in supersymmetric Yang-Mills theories
 \cite{SUSY} and for simulations of QCD with SU(2) colour at high
 densities \cite{SU2QCD}.
 In the present paper I shall consider a TSMB algorithm allowing for
 different actions in the two steps.

 For convenience of the reader let us first shortly summarize the main
 features of TSMB.
 Let us consider the case of an arbitrary number of identical fermion
 flavours $N_f$ and assume the existence of a hermitean fermion
 matrix $\tilde{Q}$, which has the determinant $\det(\tilde{Q})$
 appearing in the effective action for the gauge field after the
 integration over the fermionic variables in the path integral.
 Multi-boson algorithms are based on the representation 
 \cite{MULTIB}
\be\label{eq01}
\left|\det(\tilde{Q})\right|^{N_f} =
\left\{\det(\tilde{Q}^2)\right\}^{N_f/2}
\simeq \frac{1}{\det P_n(\tilde{Q}^2)} \ .
\ee
 Here the  polynomial $P_n$ satisfies
\be\label{eq02}
\lim_{n \to \infty} P_n(x) = x^{-N_f/2}
\ee
 in an interval $x \in [\epsilon,\lambda]$ covering the spectrum of
 $\tilde{Q}^2$.
 For the multi-boson representation of the determinant one uses
 the roots of the polynomial $r_j,\; (j=1,\ldots,n)$
\be\label{eq03}
P_n(\tilde{Q}^2) =
r_0 \prod_{j=1}^n (\tilde{Q}^2 - r_j) \ .
\ee
 Assuming that the roots occur in complex conjugate pairs, one can
 introduce the equivalent forms
\be\label{eq04}
P_n(\tilde{Q}^2)
= r_0 \prod_{j=1}^n [(\tilde{Q} \pm \mu_j)^2 + \nu_j^2]
= r_0 \prod_{j=1}^n (\tilde{Q}-\rho_j^*) (\tilde{Q}-\rho_j)
\ee
 where $r_j \equiv (\mu_j+i\nu_j)^2$ and $\rho_j \equiv \mu_j + i\nu_j$.
 With the help of complex boson (pseudofermion) fields $\Phi_{jx}$
 one can write
\begin{eqnarray}\label{eq05}
\left|\det(\tilde{Q})\right|^{N_f} &\propto&
\prod_{j=1}^n\det[(\tilde{Q}-\rho_j^*) (\tilde{Q}-\rho_j)]^{-1} 
\nonumber \\[0.5em]
&\propto&
\int [d\Phi]\; \exp\left\{ -\sum_{j=1}^n \sum_{xy}
\Phi_{jy}^+\, [(\tilde{Q}-\rho_j^*) (\tilde{Q}-\rho_j)]_{yx}\,
\Phi_{jx} \right\} \ .
\end{eqnarray}
 Since for a finite polynomial of order $n$ the approximation in
 (\ref{eq02}) is not exact, one has to extrapolate the results to
 $n\to\infty$.

 The difficulty for small fermion masses in large physical volumes is
 that the {\em condition number} $\lambda/\epsilon$ becomes very large
 ($10^4-10^6$) and very high orders $n = {\cal O}(10^3)$ are needed for
 a good approximation.
 This requires large storage and the autocorrelation of the gauge
 configurations becomes very bad since the autocorrelation length is a
 fast increasing function of $n$.
 One can achieve substantial improvements on both these problems by
 introducing a two-step polynomial approximation:
\be\label{eq06}
\lim_{n_2 \to \infty} P^{(1)}_{n_1}(x)P^{(2)}_{n_2}(x) =
x^{-N_f/2} \ , \hspace{3em}
x \in [\epsilon,\lambda] \ .
\ee
 The multi-boson representation is only used here for the first
 polynomial $P^{(1)}_{n_1}$ which provides a first crude approximation
 and hence the order $n_1$ can remain relatively low.
 The correction factor $P^{(2)}_{n_2}$ is realized in a stochastic
 {\em noisy Metropolis correction step} with a global accept-reject
 condition during the updating process.
 In order to obtain an exact algorithm one has to consider in this case
 the limit $n_2\to\infty$.

 In the two-step approximation scheme for $N_f$ flavours of fermions
 the absolute value of the determinant is represented as
\be\label{eq07}
\left|\det(\tilde{Q})\right|^{N_f} \;\simeq\;
\frac{1}{\det P^{(1)}_{n_1}(\tilde{Q}^2)
\det P^{(2)}_{n_2}(\tilde{Q}^2)} \ .
\ee
 After an update sweep over the gauge field a global accept-reject
 step is introduced in order to reach the distribution of gauge field
 variables $[U]$ corresponding to the right hand side of
 (\ref{eq07}).
 This can be done stochastically by generating a random vector
 $\eta$ according to the normalized Gaussian distribution
\be \label{eq08}
\frac{e^{-\eta^\dagger P^{(2)}_{n_2}(\tilde{Q}[U]^2)\eta}}
{\int [d\eta] e^{-\eta^\dagger P^{(2)}_{n_2}(\tilde{Q}[U]^2)\eta}}  \ ,
\ee
 and accepting the change $[U] \to [U^\prime]$ with probability
\be \label{eq09}
P_A\left( [U^\prime] \lar [U] \right) =
\min\left\{ 1,A(\eta;[U^\prime] \lar [U]) \right\} \ ,
\ee
 where
\be \label{eq10}
A(\eta;[U^\prime] \lar [U]) =
\exp\left\{-\eta^\dagger P^{(2)}_{n_2}(\tilde{Q}[U^\prime]^2)\eta 
           +\eta^\dagger P^{(2)}_{n_2}(\tilde{Q}[U]^2)\eta\right\}\ .
\ee

 The Gaussian noise vector $\eta$ can be obtained from $\eta^\prime$
 distributed according to the simple Gaussian distribution
\be \label{eq11}
\frac{e^{-\eta^{\prime\dagger}\eta^\prime}}
{\int [d\eta^\prime] e^{-\eta^{\prime\dagger}\eta^\prime}}
\ee
 by setting it equal to
\be \label{eq12}
\eta = P^{(2)}_{n_2}(\tilde{Q}[U]^2)^{-\half} \eta^\prime  \ .
\ee

 In order to obtain the inverse square root on the right hand side of
 (\ref{eq12}), one can proceed with a polynomial approximation
\be \label{eq13}
 P^{(3)}_{n_3}(x) \simeq P^{(2)}_{n_2}(x)^{-\half} \ , \hspace{1em}
x \in [0,\lambda] \ .
\ee
 This is a relatively easy approximation because
 $P^{(2)}_{n_2}(x)^{-\half}$ is not singular at $x=0$, in contrast to
 the function $x^{-N_f/2}$.
 A practical way to obtain $P^{(3)}$ is to use some approximation scheme
 for the inverse square root.
 A simple possibility is to use a Newton iteration
\be \label{eq14}
P^{(3)}_{k+1} = \half \left( P^{(3)}_k + \frac{1}{P^{(3)}_k P^{(2)}} 
\right) \ , \hspace{3em} k=0,1,2,\ldots  \ . 
\ee
 The second term on the right hand side can be evaluated by a polynomial
 approximation as for $P^{(2)}$ in (\ref{eq06}) with $N_f=0$ and
 $P^{(1)} \to P^{(3)}_k P^{(2)}$.
 The iteration in (\ref{eq14}) is fast converging and allows for an
 iterative procedure stopped by a prescribed precision.
 A starting polynomial $P^{(3)}_0$ can be obtained, for instance,
 from $P^{(2)}$ in (\ref{eq06}) with $N_f \to -\half N_f$.

 The TSMB algorithm becomes exact only in the limit of infinitely high
 polynomial order: $n_2\to\infty$ in (\ref{eq06}).
 Instead of investigating the dependence of expectation values on $n_2$
 by performing several simulations, it is better to fix some relatively
 high order $n_2$ for the simulation and perform another correction in
 the ``measurement'' of expectation values by still finer polynomials.
 This is done by {\em reweighting} the configurations.

 The reweighting for general $N_f$ is based on a polynomial
 approximation $P^{(4)}_{n_4}$ which satisfies
\be\label{eq15}
\lim_{n_4 \to \infty} P^{(1)}_{n_1}(x)P^{(2)}_{n_2}(x)P^{(4)}_{n_4}(x) =
x^{-N_f/2} \ , \hspace{3em}
x \in [\epsilon^\prime,\lambda] \ .
\ee
 The interval $[\epsilon^\prime,\lambda]$ can be chosen by convenience,
 for instance, such that $\epsilon^\prime=0,\lambda=\lambda_{max}$,
 where $\lambda_{max}$ is an absolute upper bound of the eigenvalues of
 $\tilde{Q}^2$.
 In practice, instead of $\epsilon^\prime=0$, it is more effective to
 take $\epsilon^\prime > 0$ and determine the eigenvalues below
 $\epsilon^\prime$ and the corresponding correction factors explicitly
 \cite{SUSY}.
 With properly chosen $[\epsilon^\prime,\lambda]$ the limit
 $n_4\to\infty$ is exact on an arbitrary set of gauge configurations.
 For the evaluation of $P^{(4)}_{n_4}$ one can use $n_4$-independent
 recursive relations \cite{POLYNOM}, which can be stopped by observing
 the required precision of the result.
 After reweighting, the expectation value of a quantity $\cal{O}$ is
 given by
\be\label{eq16}
\langle {\cal O} \rangle = \frac{\langle {\cal O} 
\exp{\{\eta^\dagger[1-P^{(4)}_{n_4}(\tilde{Q}^2)]\eta\}}
\rangle_{U,\eta}}
{\langle  \exp{\{\eta^\dagger[1-P^{(4)}_{n_4}(\tilde{Q}^2)]\eta\}}
\rangle_{U,\eta}} \ ,
\ee
 where $\eta$ is a simple Gaussian noise like $\eta^\prime$ in
 (\ref{eq11}).
 Here $\langle\ldots\rangle_{U,\eta}$ denotes an expectation value
 on the gauge field sequence, which is obtained in the two-step process
 described above, and on a sequence of independent $\eta$'s.

 The necessity and importance of reweighting the gauge field
 configurations depends on the condition number $\lambda/\epsilon$,
 which grows roughly as the squared inverse of the fermion mass in
 lattice units.
 For relatively heavy fermions one can choose $n_2$ high enough, such
 that the effect of reweighting is negligible compared to the
 statistical errors.
 This can be checked on a subsample of statistically independent
 configurations and then the systematic errors due to the finiteness of
 $n_2$ are under control.
 The reweighting becomes necessary only for very light fermions as,
 for example, in \cite{SU2QCD}.
 In cases if reweighting becomes important the computational work for
 obtaining the reweighting factors is comparable for the calculation of
 the inverse of $\tilde{Q}^2$ on the vectors $\eta$ in (\ref{eq16}).
 This is typically less than the off-line calculations performed,
 for instance, for determining some correlators and their matrix
 elements.
 Of course, if the reweighting has some effect, then it has to be taken
 into account in the process of estimating statistical errors
 (see \cite{ALPHA}).

\section{Variable multi-boson update step}\label{sec2}

 The TSMB algorithm is based on the fact that the change of the fermion
 determinant in the update sequence can be approximated by a
 substantially lower order polynomial than would be required for the
 final precision of the simulation.
 More generally, one can also allow for the fermion action in the
 multi-boson step to differ from the true action describing the model
 to be simulated.
 (This has been done in a special case in ref.~\cite{SMEARED} where
 the update step is performed with the pure gauge action.)
 Of course, the correction steps become in general more involved
 because one has to correct for the difference of the two actions, too.

 Let us denote the auxiliary hermitean fermion action in the multi-boson
 step in general by $\tilde{Q}_0$.
 Instead of (\ref{eq07}) the fermion determinant is now represented by
\be\label{eq17}
\left|\det(\tilde{Q})\right|^{N_f} \;\simeq\;
\frac{1}{\det \bar{P}^{(1)}_{\bar{n}_1}(\tilde{Q}_0^2)\;
\det \bar{P}^{(2)}_{\bar{n}_2}(\tilde{Q}_0^2)\;
\det P^{(2)}_{n_2}(\tilde{Q}^2)} \ .
\ee
 The main difference compared to (\ref{eq07}) is that here an
 additional polynomial ($\bar{P}^{(2)}$) appears which is needed in
 order to compensate for the use of a different action $\tilde{Q}_0$
 in the first polynomial $\bar{P}^{(1)}$.
 The polynomials appearing in (\ref{eq17}) have to satisfy, with
 $R^{(..)} \simeq 1$ and $\bar{R}^{(..)} \simeq 1$, the relations
\be\label{eq18}
x^{-N_f/2} = \frac{1}{\bar{P}^{(1)}(x) \bar{R}^{(1)}(x)}
= \frac{1}{P^{(2)}(x) R^{(2)}(x)} \ , \hspace{3em}
\bar{P}^{(1)}(x) \bar{P}^{(2)}(x) = \bar{R}^{(12)}(x) \ .
\ee
 Here $R^{(2)}(x) \simeq 1$ and $\bar{R}^{(12)}(x) \simeq 1$ have to
 be much better approximations than $\bar{R}^{(1)}(x) \simeq 1$.

 The factor $(\det\bar{P}^{(1)})^{-1}$ in (\ref{eq17}) is generated
 by the multi-boson update step using the action $\tilde{Q}_0$.
 The remaining factor, which has to be reproduced by the noisy
 Metropolis step, can be written, for instance, as
\be\label{eq19}
\frac{1}{\det \bar{P}^{(2)}(\tilde{Q}_0^2)\;\det P^{(2)}(\tilde{Q}^2)}
= \frac{1}{\det\left( \bar{P}^{(2)}(\tilde{Q}_0^2)^\half\;
 \det P^{(2)}(\tilde{Q}^2)\;
 \bar{P}^{(2)}(\tilde{Q}_0^2)^\half\right)} \ . 
\ee
 Note that here and in what follows the r\^oles of $P^{(2)}$ and
 $\bar{P}^{(2)}$ can also be exchanged.

 Using the form in (\ref{eq19}) one can write the detailed balance
 condition for the acceptance probability $P_A$ in (\ref{eq09}) as
\begin{eqnarray}\label{eq20}
\frac{P_A\left( [U^\prime] \lar [U] \right)}
{P_A\left( [U] \lar [U^\prime] \right)}
&=& \frac{\det\left(
\bar{P}^{(2)}[U]^\half P^{(2)}[U] \bar{P}^{(2)}[U]^\half \right)}
{\det\left( \bar{P}^{(2)}[U^\prime]^\half P^{(2)}[U^\prime]
 \bar{P}^{(2)}[U^\prime]^\half \right)}
\nonumber \\[0.5em]
&=& \frac{ \int [d\eta] \exp\left( -\eta^\dagger
 \bar{P}^{(2)}[U^\prime]^\half P^{(2)}[U^\prime]
 \bar{P}^{(2)}[U^\prime]^\half \eta \right)}
{ \int [d\eta] \exp\left( -\eta^\dagger
 \bar{P}^{(2)}[U]^\half P^{(2)}[U] \bar{P}^{(2)}[U]^\half
 \eta\right)} \ .
\end{eqnarray}
 Here the gauge field dependence of the polynomials is displayed,
 which reflects the dependence of the fermion actions on the gauge
 field.
 The detailed balance condition can be satisfied similarly to
 (\ref{eq09}) with
\be \label{eq21}
A(\eta;[U^\prime] \lar [U]) =
\exp\left\{- \eta^\dagger
\bar{P}^{(2)}[U^\prime]^\half P^{(2)}[U^\prime]
\bar{P}^{(2)}[U^\prime]^\half \eta 
+ \eta^\dagger \bar{P}^{(2)}[U]^\half P^{(2)}[U]
\bar{P}^{(2)}[U]^\half \eta\right\}\ .
\ee
 The required Gaussian can now be obtained from a simple one in
 (\ref{eq11}) by
\be \label{eq22}
\eta = \bar{P}^{(2)}[U]^{-\half} P^{(2)}[U]^{-\half} \eta^\prime  \ .
\ee
 For the evaluation of (\ref{eq21}) we need the polynomial
 approximations
\be \label{eq23}
P^{(3)}_{n_3}(x) \simeq P^{(2)}_{n_2}(x)^{-\half} \ , \hspace{2em}
\bar{P}^{(3)}_{\bar{n}_3}(x) \simeq 
\bar{P}^{(2)}_{\bar{n}_2}(x)^{-\half} \ , 
\hspace{2em} x \in [0,\lambda] \ ,
\ee
 and
\be \label{eq24}
P^{(-3)}_{n_{-3}}(x) \simeq P^{(3)}_{n_3}(x)^{-1} \simeq
P^{(2)}_{n_2}(x)^\half \ , \hspace{2em}
\bar{P}^{(-3)}_{\bar{n}_{-3}}(x) \simeq 
\bar{P}^{(3)}_{\bar{n}_3}(x)^{-1} \simeq
\bar{P}^{(2)}_{\bar{n}_2}(x)^\half \ , 
\hspace{2em} x \in [0,\lambda] \ .
\ee

 The measurement correction can be performed in an analogous manner
 as before.
 For this we need the polynomial approximations $P^{(4)}(x)$ and 
 $\bar{P}^{(4)}(x)$ corresponding to (\ref{eq18}):
\be\label{eq25}
x^{-N_f/2} = \frac{1}{P^{(2)}(x) P^{(4)}(x) R^{(24)}(x)} \ , 
\hspace{2em}
\bar{P}^{(1)}(x) \bar{P}^{(2)}(x) \bar{P}^{(4)}(x) = 
\bar{R}^{(124)}(x) \ .
\ee
 The final precision is given now by the quality of the approximations
 $R^{(24)}(x) \simeq 1$ and $\bar{R}^{(124)}(x) \simeq 1$.

\section{An application}\label{sec3}

 Apart from the doubling of the number of polynomials in the noisy
 Metropolis correction step the present algorithm is entirely analogous
 to normal TSMB.
 As a first application let us here consider the case where the two
 hermitean actions $\tilde{Q}_0$ and $\tilde{Q}$ are the same Wilson
 fermion action, except for the values of the parameters $\beta$
 (the gauge coupling) and $\kappa$ (the hopping parameter).

 Let us denote the parameters in the multi-boson step by
 ($\beta_0,\kappa_0$) and in the noisy Metropolis step by
 ($\beta,\kappa$).
 An interesting question is how for fixed parameters ($\beta,\kappa$)
 the change of ($\beta_0,\kappa_0$) does influence the behaviour
 of the algorithm.
 To see this I performed a test run with two flavours of Wilson
 fermions on an $8^3 \cdot 16$ lattice at parameters
 ($\beta=5.28,\kappa=0.160$).
 (This corresponds to a point on the $N_t=4$ cross-over line of
 \cite{THERMODYN}.)
 The main polynomial parameters were as follows:
 $[\epsilon,\,\lambda]=[0.00875,\,2.8],\,n_1=24,\,n_2=70$.

\begin{figure}[tbh]
\vspace*{20mm}
\begin{flushleft}
\epsfig{file=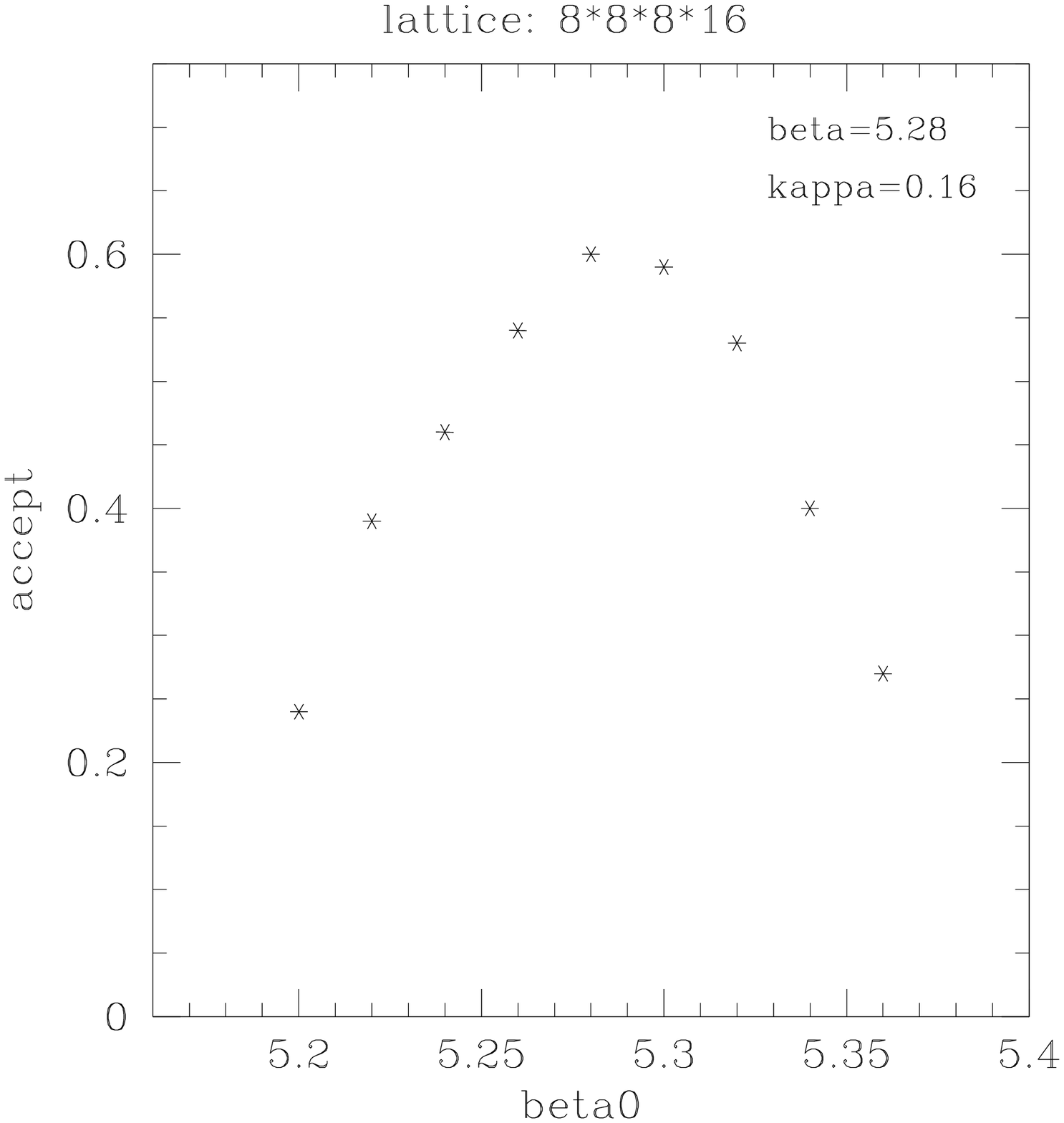,
        width=8.0cm,height=8.0cm,
        angle=0}
\end{flushleft}
\vspace*{-90mm}
\begin{flushright}
\epsfig{file=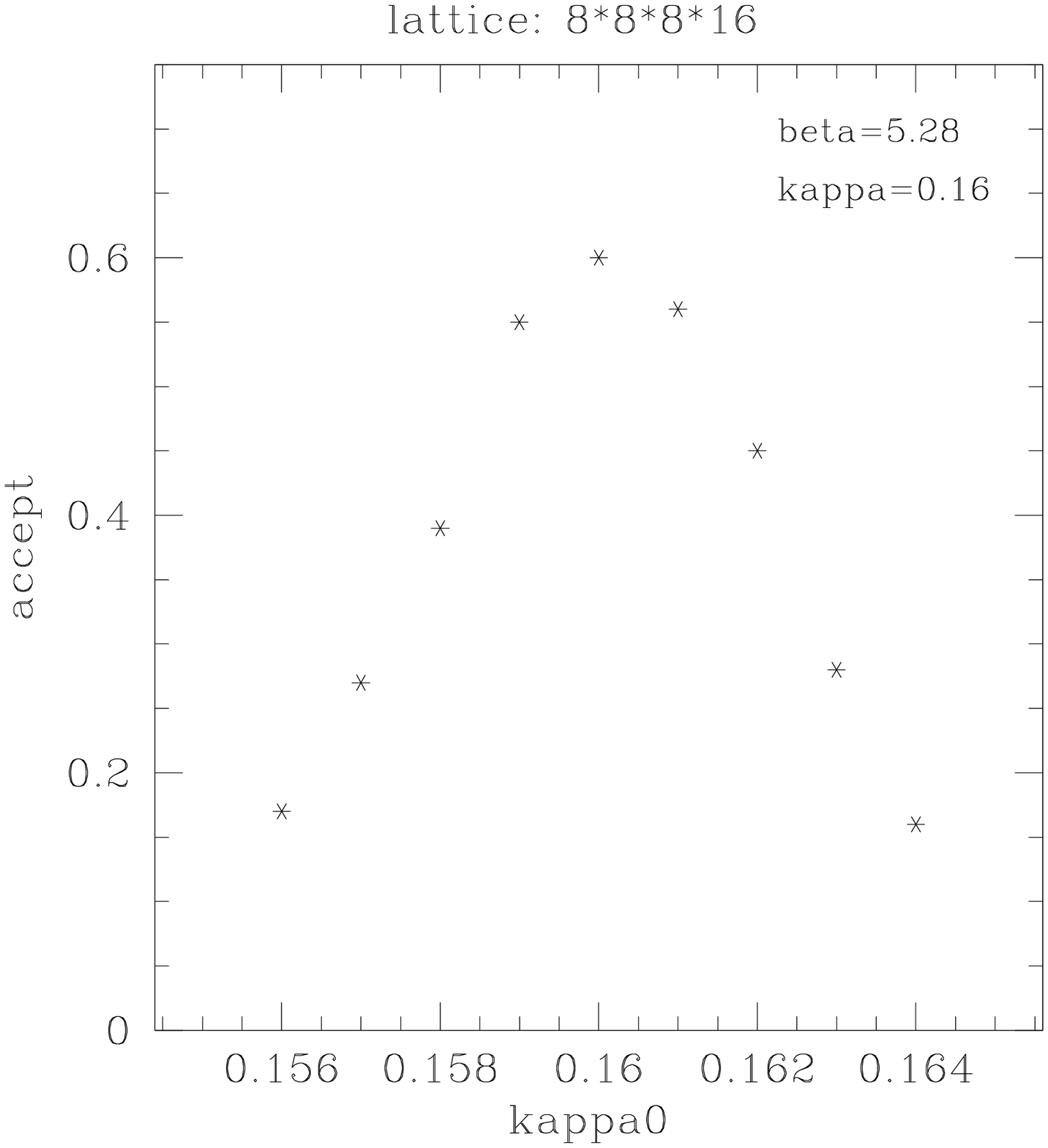,
        width=8.0cm,height=8.0cm,
        angle=0}
\end{flushright}
\vspace*{-6mm}
\begin{center}
\parbox{15cm}{\caption{\label{fig01}
 The average acceptance of the noisy Metropolis step at
 ($\beta=5.28,\,\kappa=0.16$).
 On left (right): changing $\beta_0$ ($\kappa_0$) in the multi-boson
 update step.
}}
\end{center}
\end{figure}

\begin{figure}[tbh]
\vspace*{-15mm}
\begin{flushleft}
\epsfig{file=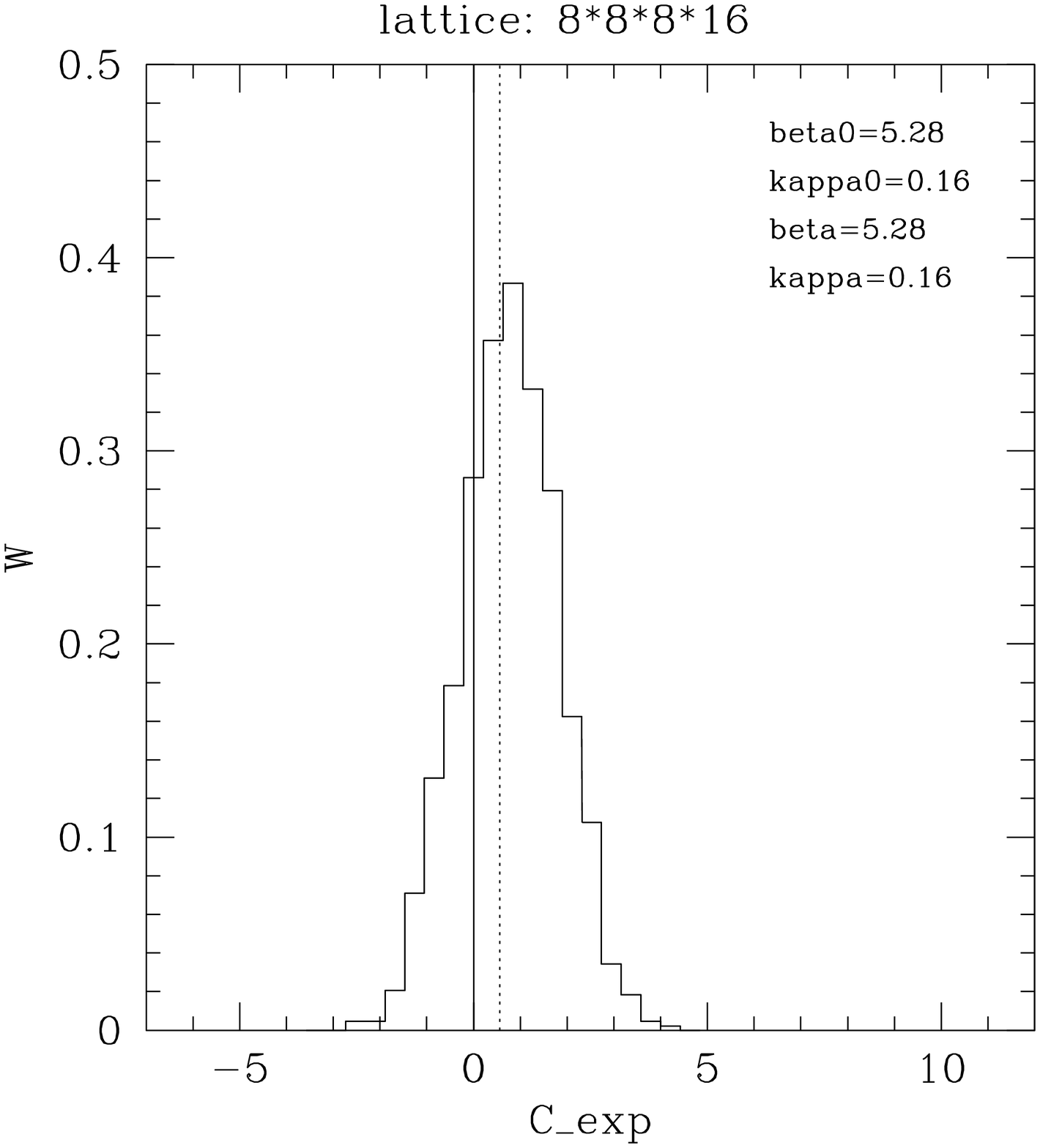,
        width=8.0cm,height=8.0cm,
        angle=0}
\end{flushleft}
\vspace*{-90mm}
\begin{flushright}
\epsfig{file=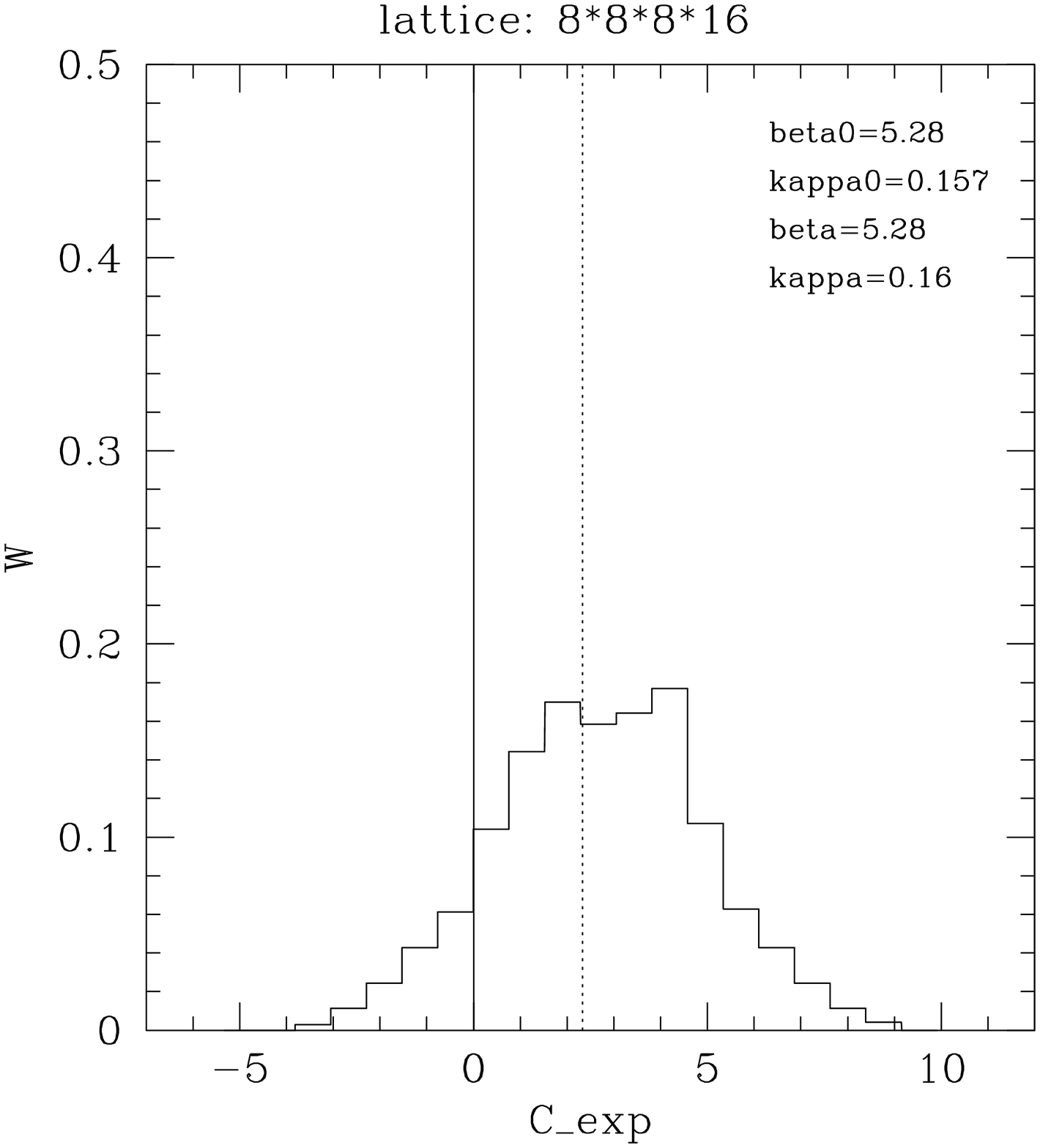,
        width=8.0cm,height=8.0cm,
        angle=0}
\end{flushright}
\vspace*{-6mm}
\begin{center}
\parbox{15cm}{\caption{\label{fig02}
 The distribution of the exponent in the noisy Metropolis step at 
 ($\beta=5.28,\,\kappa=0.16$).
 The parameters in the multi-boson update step: on the left (right)
 $\beta_0=5.28,\,\kappa_0=0.160$ ($\beta_0=5.28,\,\kappa_0=0.157$).
 The vertical lines show the mean value of the distributions (and
 zero).
}}
\end{center}
\end{figure}

 The change of the acceptance rate in the noisy Metropolis step for
 changing $\beta_0$ resp. $\kappa_0$ is shown by figure \ref{fig01}.
 As one can see, the acceptance remains quite good in a relatively
 broad range of parameters.
 This is partly due to the fact that the distribution of the exponent
 in the noisy Metropolis step becomes substantially broader when
 the multi-boson update parameters ($\beta_0,\kappa_0$) differ from
 ($\beta,\kappa$) (see figure \ref{fig02}).

 Changing ($\beta_0,\,\kappa_0$) can be used to introduce more
 randomness in the update process.
 In fact, since the final distribution of gauge configurations does not
 depend on the parameters of the multi-boson update, during a run one
 can randomly change ($\beta_0,\kappa_0$).
 This improves the autocorrelations.
 As an example, the measured integrated autocorrelation of the plaquette
 in the original TSMB run with the above polynomial parameters is
 $\tau_{int}^{plaq} = 2.9(3) \cdot 10^5 MVM$
 (Matrix-Vector Multiplications by the even-odd preconditioned
 Wilson-Dirac matrix).
 This is decreased to $\tau_{int}^{plaq} = 1.2(3) \cdot 10^5 MVM$ if
 (for fixed $\kappa_0=\kappa$) $\beta_0$ is changed randomly during the
 gauge update in the interval $\beta_0 \in [5.16,\, 5.40]$.
 For counting the number of MVM's in an update cycle the following
 approximate formula is used:
\be\label{eq26}
N_{MVM} \simeq 6(n_1 N_B + N_G) + 
2N_C (n_2+n_3+\bar{n}_3+\bar{n}_{-3}) \ .
\ee
 Here $N_B$ is the number of boson field updates, $N_G$ the gauge field
 updates and $N_C$ the number of noisy Metropolis accept-reject steps.
 The autocorrelations were determined in independent runs which were at
 least as long as hundred times the measured autocorrelation.
 In these test runs with relatively heavy quarks the polynomial order
 $n_2$ is high enough and the effect of reweighting in (\ref{eq16})
 is much smaller than the statistical errors.
 Therefore, the autocorrelations were determined without taking into
 account the reweighting factors.

 The decrease of the plaquette autocorrelation as an effect of updating
 with a ``wrong'' action is at the first sight against intuition.
 This unexpected effect can be understood as due to partly compensating
 the dominance of the bosonic contributions against the pure gauge
 contributions in the effective gauge action.
 As a consequence of the gauge field part in (\ref{eq21}), the
 fluctuation of the gauge fields within the natural band of fluctuations
 is intensified.
 At the same time there is some decrease of the acceptance, too, but
 in total the effect of the amplified fluctuations is more important.

 An interesting question is the volume dependence of the effect of
 action variations in the update.
 In general, the allowed action changes may be expected to decrease
 in larger lattice volumes.
 However, one has to observe that the range with good acceptance
 in figure \ref{fig01} is much larger than the range which would be
 allowed for entirely updating with different parameters and
 performing the necessary reweighting afterwards on the equilibrium
 configurations.
 The main reason is that the distance between two noisy Metropolis
 steps is typically less than 1\% of the plaquette autocorrelation
 distance.
 In addition, on larger volumes the number of boson fields $n_1$ is
 also larger and the dominance of the bosonic part in the effective
 gauge action is stronger. 
 Therefore the question of the volume dependence is not easy to
 answer and can be best done by performing actual test runs on
 larger volumes \cite{QCD3}.

 Another interesting possibility is to apply TSMB for variable actions
 in numerical simulations with improved fermion actions.
 Using a simplified version of the particular improved action in the
 multi-boson update step facilitates the implementation of TSMB
 algorithms and may also improve the autocorrelations.

\vspace*{1em}\noindent
{\large\bf Acknowledgement}

\noindent
 I thank Hartmut Wittig and Claus Gebert for a critical reading of the
 manuscript.



\newpage\vspace*{1em}
\begin{center}
{\large\bf Figures}
\end{center}

\end{document}